\documentclass[sigconf]{acmart}

\AtBeginDocument{%
  }

\usepackage{graphicx}
\usepackage{booktabs}
\usepackage{multirow}
\usepackage{enumitem}
\usepackage{subcaption}

\graphicspath{{figures/}}

\setcopyright{acmlicensed}
\copyrightyear{2026}
\acmYear{2026}
\setcopyright{cc}
\setcctype{by}
\acmConference[ETRA '26]{2026 Symposium on Eye Tracking Research and Applications}{June 01--04, 2026}{Marrakesh, Morocco}
\acmBooktitle{2026 Symposium on Eye Tracking Research and Applications (ETRA '26), June 01--04, 2026, Marrakesh, Morocco}
\acmDOI{10.1145/3797246.3805721}
\acmISBN{979-8-4007-2519-7/2026/06}

\begin{document}

\title{Analyzing Visual Attention Patterns During Band Rehearsal with Mobile Eye Tracking}


\author{Arvind Srinivasan}
\affiliation{%
  \institution{Aarhus University}
  \city{Aarhus N}
  \country{Denmark}}
\email{arvind@cs.au.dk}
\orcid{0000-0002-3409-6077}

\author{Tobias Rau}
\email{tobias.rau@visus.uni-stuttgart.de}
\orcid{0000-0002-3310-9163}
\affiliation{%
  \institution{University of Stuttgart}
  \city{Stuttgart}
  \country{Germany}}

\author{Michael Sedlmair}
\email{michael.sedlmair@visus.uni-stuttgart.de}
\orcid{0000-0001-7048-9292}
\affiliation{%
  \institution{University of Stuttgart}
  \city{Stuttgart}
  \country{Germany}}

\renewcommand{\shortauthors}{Srinivasan et al.}

\begin{abstract}
Visual attention is central to ensemble coordination, yet how musicians allocate gaze during naturalistic rehearsal remains poorly understood. We present a pilot study using mobile eye tracking to examine gaze behaviour in a four-member band across three songs, each practiced twice. Musicians wore Pupil Labs Neon eye trackers, and YOLOv8-assisted scene annotations mapped fixations to ensemble members and objects in view. Analyzing fixation matrices, transition matrices, temporal scarf plots, and dwell--transition correlations, we uncover a hub-and-spoke attention topology: the session leader was the dominant gaze target for all members, while the learning guitarist concentrated up to 97\% of interpersonal dwell on this single reference. Between attempts,
gaze transitions decreased by up to 65\% on average for unfamiliar material (up to 82\% for individual participants) as scanning stabilized. Scarf plots reveal how teaching breakdowns fragment attention and uninterrupted runs consolidate it. Post-session participant reflections align with the quantitative patterns, and we discuss implications for gaze-aware tools in ensemble pedagogy.
\end{abstract}

\ccsdesc[500]{Human-centered computing~HCI design and evaluation methods}
\ccsdesc[500]{Human-centered computing~Empirical studies in HCI}
\ccsdesc[300]{Human-centered computing~Collaborative and social computing theory, concepts and paradigms}

\keywords{eye tracking, visual attention, collaborative music performance, gaze coordination, ensemble rehearsal, mobile eye tracking}


\begin{teaserfigure}
    \centering
    \includegraphics[width=0.8\textwidth]{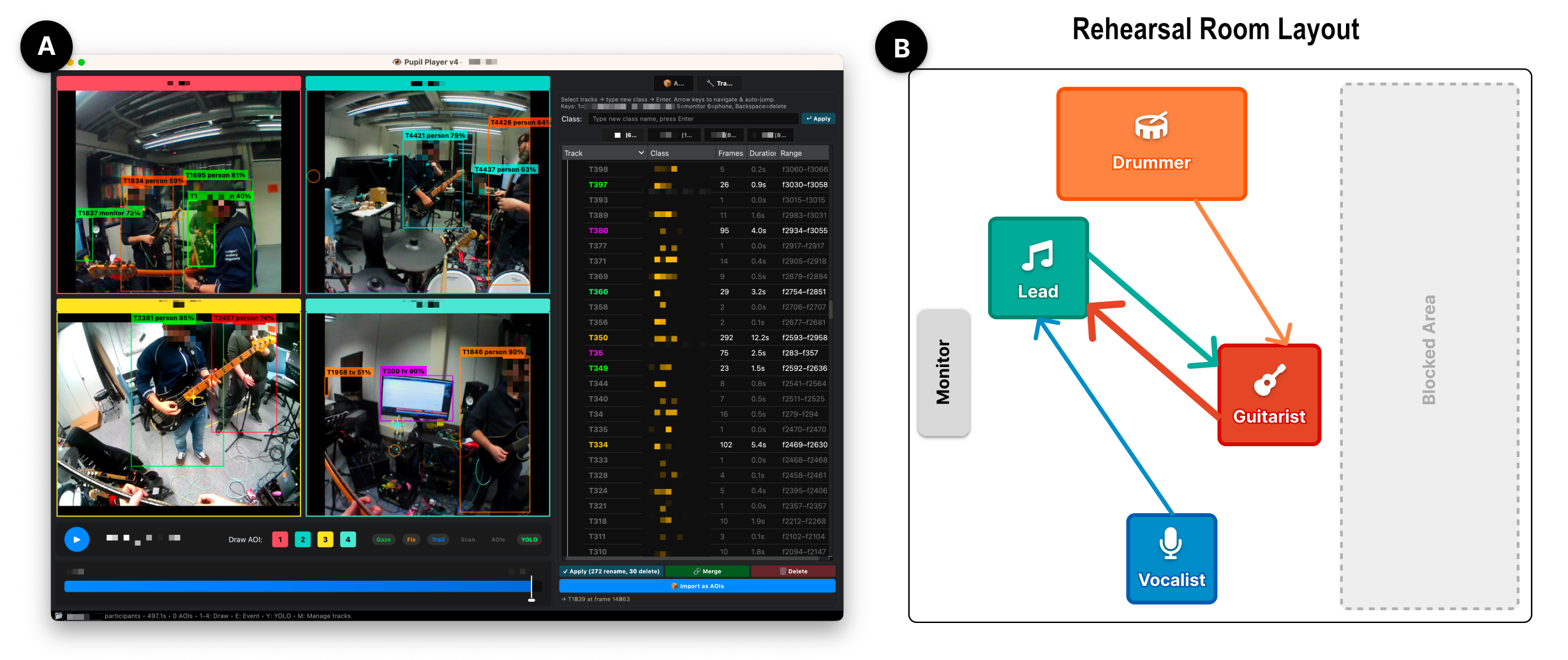}
    \caption{\textbf{Overview of our rehearsal study and analysis setup}. (A) Multi-camera recording interface with real-time person detection, bounding boxes, and track timelines used to annotate attention and interaction events. (B) Spatial layout of the rehearsal room showing participant positions—Vocalist, Drummer, Guitarist, and Lead—along with monitor placement and a blocked area. Colored arrows indicate observed attention and interaction flows between band members.}
    \Description{Two-panel figure. Panel A shows a multi-camera rehearsal recording interface with four video feeds and overlaid detection boxes and track timelines. Panel B shows a diagram of the rehearsal room layout with labeled participant positions (Vocalist, Drummer, Guitarist, Lead), a monitor, and a blocked area, with arrows indicating attention and interaction directions between members.}
  \label{fig:teaser}
\end{teaserfigure}

\maketitle

\section{Introduction}

Music ensemble performance requires real-time coordination with millisecond precision, typically without verbal communication during play~\cite{keller2014rhythm}. Visual attention enables musicians to anticipate co-performers' actions, synchronize timing, and recover from errors~\cite{bishop2019gaze}, yet how musicians distribute gaze during naturalistic rehearsal remains poorly understood. Prior eye-tracking work has focused on sight-reading~\cite{drai2012effect} or used stationary trackers in dyadic setups~\cite{bishop2019moving, vandemoortele2018gaze}, leaving multi-performer gaze dynamics during band practice largely unexamined.

We present an exploratory pilot study addressing the question: \emph{How do musicians distribute their visual attention during band rehearsal, and how do these patterns reflect collaborative roles and learning?} We equipped four members of an amateur band with mobile eye trackers (Pupil Labs Neon) during a naturalistic rehearsal spanning three songs, each practiced twice. We analyze gaze data through \textit{fixation matrices} (proportional dwell time per target) and \textit{transition matrices} (gaze shift frequencies), complemented by temporal scarf plots, dwell--transition correlation analysis, and participant reflections. Our contributions are:
\begin{enumerate}[noitemsep,partopsep=0pt,topsep=0pt,parsep=0pt]
    \item Fixation and transition matrices across 3~songs~$\times$~2~attempts reveal a hub-and-spoke attention topology centered on the session leader, with transitions decreasing up to 65\% for unfamiliar material.
    \item Temporal scarf plots aligned with rehearsal events show how teaching breakdowns fragment attention and uninterrupted runs consolidate it.
    \item Dwell--transition correlation analysis generates testable hypotheses about attentional narrowing, illustrated by the learning guitarist and triangulated with post-session reflections.
\end{enumerate}

\section{Background}

Eye tracking in music has focused primarily on sight-reading~\cite{drai2012effect}, but ensemble musicians also visually monitor co-performers for coordination cues~\cite{bishop2019gaze}. Body movement in string quartets becomes more predictable during ensemble versus solo performance~\cite{glowinski2013movements}, and motion in duos serves communicative functions~\cite{bishop2019moving}. Keller et al.\ reviewed neurophysiological mechanisms underlying joint musical action, emphasizing anticipatory processes partly supported by visual monitoring~\cite{keller2014rhythm}. Despite these insights, mobile eye tracking of multi-performer gaze during naturalistic rehearsal remains rare.
Tomasello's framework~\cite{tomasello1995joint} distinguishes \emph{shared attention} (attending to the same target) from \emph{mutual attention} (reciprocated awareness), both supporting coordination through action prediction~\cite{brennan2008coordinating}. Familiarity with a co-performer's part affects coordination quality~\cite{ragert2013synchronization}, suggesting visual monitoring compensates for incomplete musical knowledge.

Real-time mutual gaze perception enhances collaborative learning~\cite{schneider2013real}, and temporal coupling of eye movements predicts discourse comprehension~\cite{richardson2005looking}. Krejtz et al.'s entropy-based transition analysis~\cite{krejtz2014entropy}---higher entropy indicating distributed scanning---provides the analytical foundation for our matrices. Under high cognitive load, attention narrows toward salient cues---attentional tunneling~\cite{easterbrook1959effect}---documented in aviation~\cite{wickens2009attentional}, driving~\cite{dirkin1983cognitive}, and cockpit environments~\cite{dehais2014failure}. In ensembles, tunneling may manifest as a struggling performer locking gaze onto a single reference.
\section{Method}

\paragraph{Participants}
Four amateur musicians from an established band participated; all had played together for at least one year but varied in familiarity with the recorded songs. 

\paragraph{Apparatus}
Each participant wore a Pupil Labs Neon mobile eye tracker (binocular, $\nu = 200$\,Hz gaze sample frequency; scene camera at 30\,fps, 1600$\times$1200\,px) with deep-learning-based calibration-free gaze estimation. A single-point validation was performed per session. Recordings were synchronized using Pupil Labs' time-sync protocol.

\paragraph{Procedure} The recording took place during a regular rehearsal. Three rock/metal songs were practiced, each performed twice (Attempt~1, A1; Attempt~2, A2): \textbf{S1}~(familiar, well-rehearsed), \textbf{S2}~(newer material with active teaching by the Lead), and \textbf{S3}~(moderately familiar). Between attempts the Lead provided feedback and directed re-starts, with the Guitarist as primary correction recipient. Mean duration was approximately 7.9~minutes per session per participant. Because songs were always practiced in the same fixed order, song-specific effects cannot be separated from session-order effects such as fatigue or warm-up.

\subsection{Areas of Interest (AOIs)}

AOIs comprised each participant, the \emph{monitor} (lyrics/tablature), and self-gaze (own hands/instrument). Scene videos were processed with a YOLOv8 detector fine-tuned on hand-labelled data; per-frame detections were mapped to gaze coordinates to assign fixations to AOIs.

The fine-tuned YOLOv8n detector achieved precision 0.75, recall 0.79, and mAP@50 of 0.79 (mAP@50-95 = 0.59) after three training epochs on hand-labelled frames. Monitor and phone classes were occasionally confused with background during inspection, corrected in a manual annotation pass.

\subsection{Data Processing}

Raw gaze data were processed using Pupil Labs' fixation and saccade detection. We computed three representations by mapping their targets of interest (band members) as follows:

\paragraph{Fixation (Dwell-Time) Matrix.}
For each participant~$A$, the proportion of total dwell time spent on each target ~$j$, expressed as a percentage:
\[
D_{Aj} = \frac{\sum_{k=1}^{n} d_k \cdot \mathbf{1}[\text{AOI}(f_k) = \alpha]}{\sum_{k=1}^{n} d_k} \times 100 \qquad ,
\]
where $n$ is the total number of gaze samples for participant~$A$, $f_k$ is the $k^{th}$ gaze sample, $\text{AOI}(f_k)$ maps each sample's coordinates to a target, and $d_k = 5\ ms$ is the duration of each gaze sample, at 200Hz sampling rate.

\paragraph{Transition Matrix.}
For each participant, the count of consecutive fixation pairs where gaze shifted from AOI~$\alpha$ to AOI~$\beta$:
\[
T_{\alpha\beta} = \sum_k^n \mathbf{1}[\text{AOI}(f_k) = \alpha \;\wedge\; \text{AOI}(f_{k+1}) = \beta] \qquad .
\]

In total, the dataset comprises 20\,809 fixation events, 20\,798 saccades, 145 dwell-time records and 320 transition records across all sessions.
The ``Other'' category---fixations not classified as directed toward an ensemble member or the monitor, including phone detections and unidentified background objects---accounted for 38\% of total dwell time on average.

\paragraph{Normalization conventions.}
All dwell percentages are \emph{classified-target dwell}: the denominator is restricted to the five named AOIs (Vocalist, Drummer, Guitarist, Lead, Monitor), with Other excluded and renormalized to 100\%. For interpersonal comparisons we additionally report \emph{interpersonal dwell}, further restricting the denominator to person-directed fixation only (monitor excluded). We refer to participants by role: \textbf{Vocalist}, \textbf{Drummer}, \textbf{Guitarist}, and \textbf{Lead}. The spatial layout (Figure~\ref{fig:teaser}(B)) places the Lead centrally with sightlines to all members, while the Drummer sits at the periphery; these positional constraints should be considered when interpreting dwell asymmetries. Because session durations varied (mean 5.0--12.7~minutes across conditions), we report both raw transition counts (to preserve interpretability of the scarf-plot and matrix visualizations) and rate-normalized values (transitions per minute) where relevant.

\section{Results}

\subsection{Dwell-Time Distribution}

\begin{figure*}[t]
    \centering
    \begin{subfigure}[t]{0.49\textwidth}
        \centering
        \includegraphics[width=\textwidth]{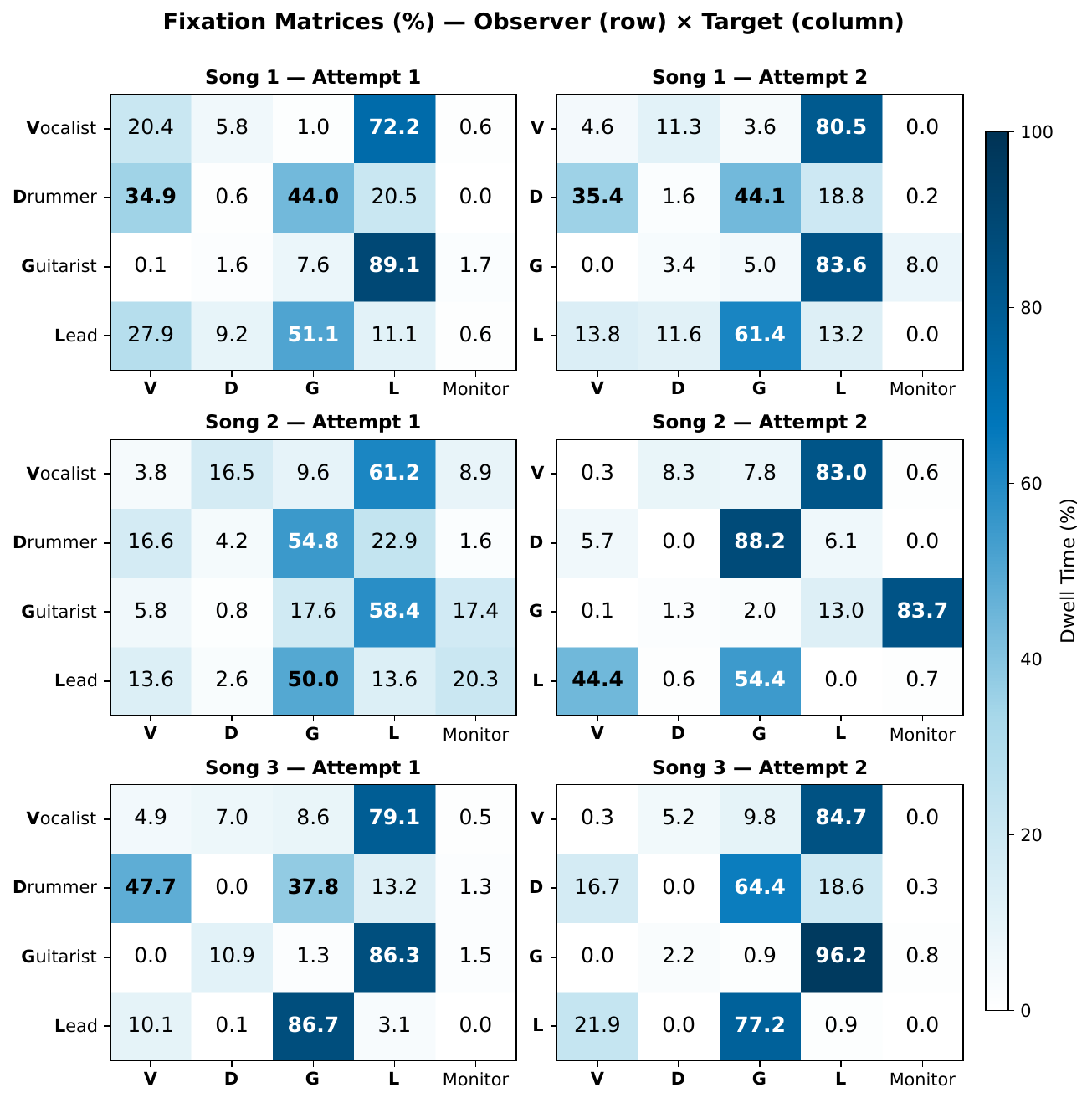}
        \caption{Dwell-time matrices (\%) across all six sessions.}
        \Description{Six heatmap panels arranged in a 2x3 grid (3 songs, 2 attempts each). Each panel is a 4x5 matrix showing classified-target dwell percentages from observer (row) to target (column: V, D, G, L, Monitor). The Lead column is consistently darkest for most observers, while the Drummer column remains light across all panels.}
        \label{fig:dwell_heatmaps}
    \end{subfigure}
    \hfill
    \begin{subfigure}[t]{0.49\textwidth}
        \centering
        \includegraphics[width=\textwidth]{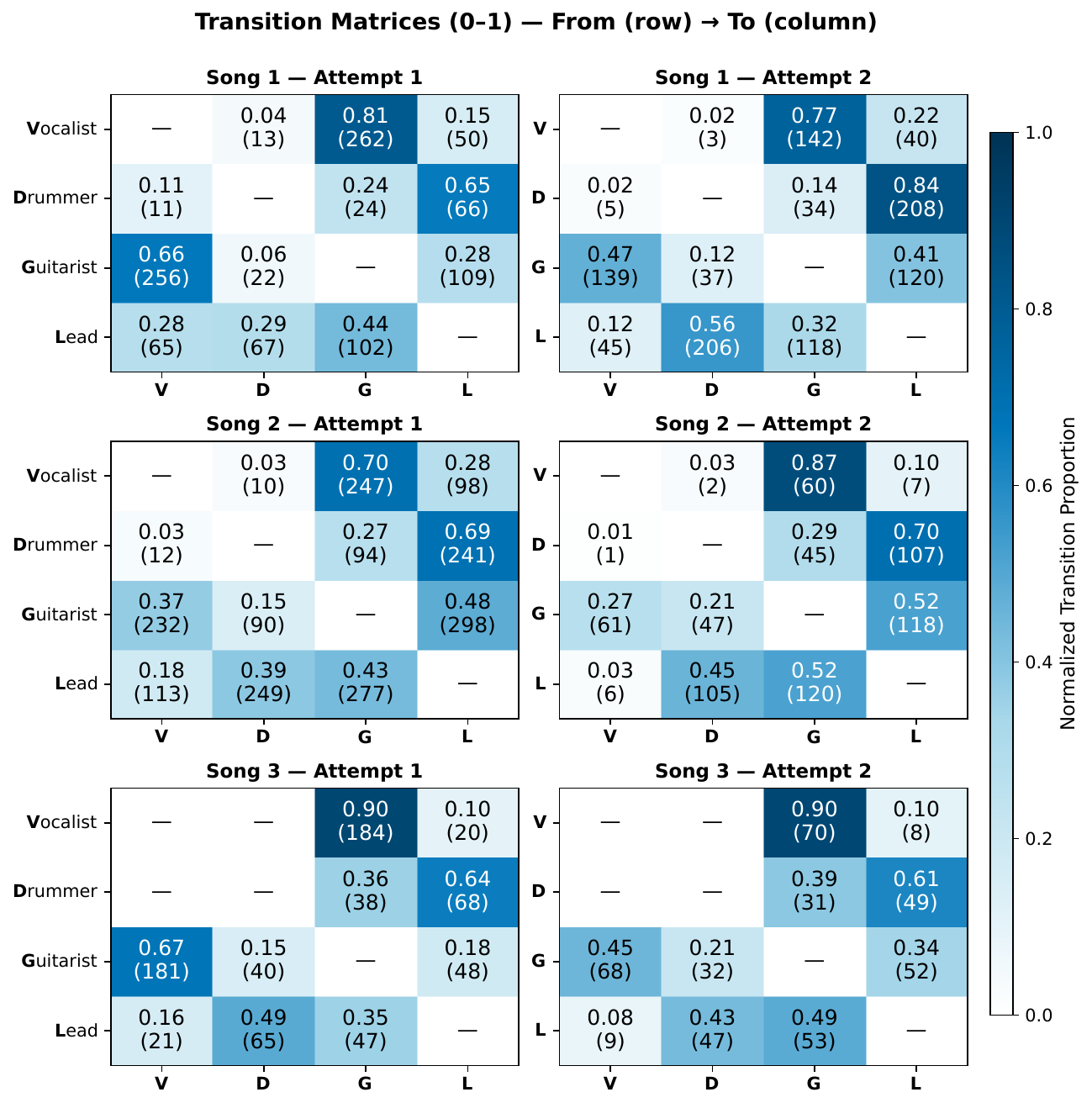}
        \caption{Normalized gaze transition matrices across all sessions. }
        \Description{Six normalized heatmap panels in a 2x3 grid. Each panel shows a 4x4 transition probability matrix with participant roles as both rows and columns. Darker cells indicate more frequent transitions. The Lead row/column consistently shows the darkest values, indicating it is the primary transition partner for most members.}
        \label{fig:transition_heatmaps}
    \end{subfigure}
    \caption{In~(a), each cell shows the proportion of classified-target dwell (Other and Phone excluded, renormalized) from observer (row) to target (column). In~(b), cells show the proportion of outgoing transitions directed from row to column. Both views confirm a hub-and-spoke topology centered on the Lead, with the Guitarist--Lead axis as the dominant coupling.}
    \label{fig:heatmaps_combined}
\end{figure*}

Figure~\ref{fig:dwell_heatmaps} presents the dwell-time matrices for all six sessions.
The dominant pattern is the concentration of attention on the Lead: the Vocalist directed 61--85\% and the Guitarist 58--96\% of dwell toward the Lead. The Lead's own gaze was more distributed, with the Guitarist as primary target in all six sessions (50--87\% of dwell). The Drummer received only 5.5\% of other members' dwell on average; while partly positional (Figure~\ref{fig:teaser}(B)), the data from YOLO detections show the Drummer appeared at moderate rates in scene cameras yet was rarely fixated. 

Under interpersonal normalization, the Guitarist allocated 71--97\% of person-directed attention to the Lead (peaking at 96.9\% in S3~A2), while the Lead devoted 51--87\% to the Guitarist. An important counterpoint emerged in S2~A2: following teaching in A1, the Guitarist's dwell on the Lead dropped from 58.4\% to 13.0\% while monitor usage surged from 17.4\% to 83.7\%. Yet interpersonal concentration on the Lead \emph{increased} (70.7\% to 79.6\%), because the shift was from person-watching to reference-reading, not from one person to another.

\subsection{Gaze Transitions}

The transition matrices (Figure~\ref{fig:transition_heatmaps}) capture gaze shifts between AOIs, with counts ranging from 41 (Guitarist, S3~A2) to 1\,015 (Vocalist, S2~A1). Mean counts per song show that the largest decrease occurred in S2 ($-65\%$), where A1 contained an extended teaching session (mean 12.7~min) and A2 a cleaner run (8.6~min). S3 showed $-43\%$; S1, the familiar song, was stable ($+5\%$). 

These decreases partly reflect shorter A2 durations (e.g., S2: 12.7 to 8.6~min on average), but the 65\% transition drop substantially exceeds the 32\% duration decrease, confirming genuine stabilization. The sharpest S2 drops were the Drummer (--82\%) and Guitarist (--71\%), though all members showed reductions, indicating the shift from stop-start to continuous playing stabilized visual behaviour group-wide.

\subsection{Dwell--Transition Correlation}

\begin{figure*}[t]
    \centering
    \includegraphics[width=0.9\linewidth]{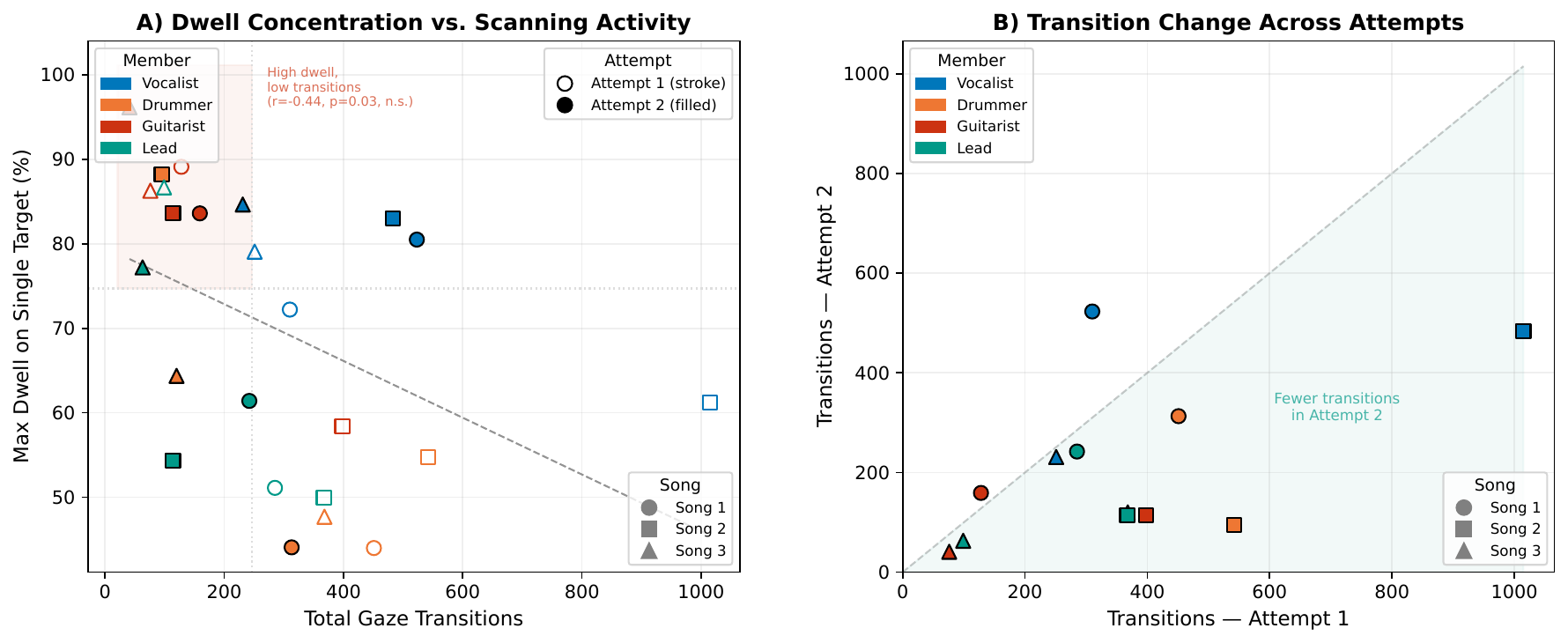}
    \caption{(A)~Relationship between maximum classified-target dwell (\%) and total transition count across all 24 recordings ($r=-0.44$, $p=0.03$). The Guitarist clusters in the high-dwell, low-transition region consistent with attentional narrowing. The Vocalist and Lead occupy more distributed positions. (B)~A1 versus A2 transition counts per participant; points below the diagonal indicate reduced scanning in the second attempt, with the steepest drop in S2.}
    \Description{Two-panel figure. Panel A: scatter plot with total gaze transitions on x-axis and max single-target dwell percentage on y-axis. Points are color-coded by participant role with shape encoding song. Guitarist data points cluster in the upper-left (low transitions, high dwell), while other participants scatter more broadly. Panel B: scatter plot of Attempt 1 versus Attempt 2 transition counts; most points fall below the diagonal, indicating fewer transitions in Attempt 2.}
    \label{fig:correlation}
\end{figure*}

Figure~\ref{fig:correlation}(A) plots maximum classified-target dwell~(\%) against total transition count across all 24 sessions. A moderate negative correlation ($r=-0.44$, $p=0.03$) indicates that higher dwell concentration on a single target is associated with fewer gaze transitions. The Guitarist consistently occupies the high-dwell, low-transition region, while the Vocalist and Lead are more distributed---consistent with attentional narrowing, though the small sample precludes strong causal claims. The Guitarist appears in this quadrant in five of six sessions, while the Drummer often showed the opposite pattern---distributed attention with frequent transitions---fitting a timekeeper's scanning role. One outlier---the Vocalist in S2~A1 (moderate dwell, 1015 transitions)---reflects the fragmented teaching session. Figure~\ref{fig:correlation}(B) plots A1 against A2 transition counts per participant; most points fall below the diagonal, confirming reduced scanning in A2, with the largest drop in S2.

\subsection{Changes Between Attempts}

\begin{figure*}[t]
    \centering
    \begin{subfigure}[t]{0.48\textwidth}
        \centering
        \includegraphics[width=\textwidth]{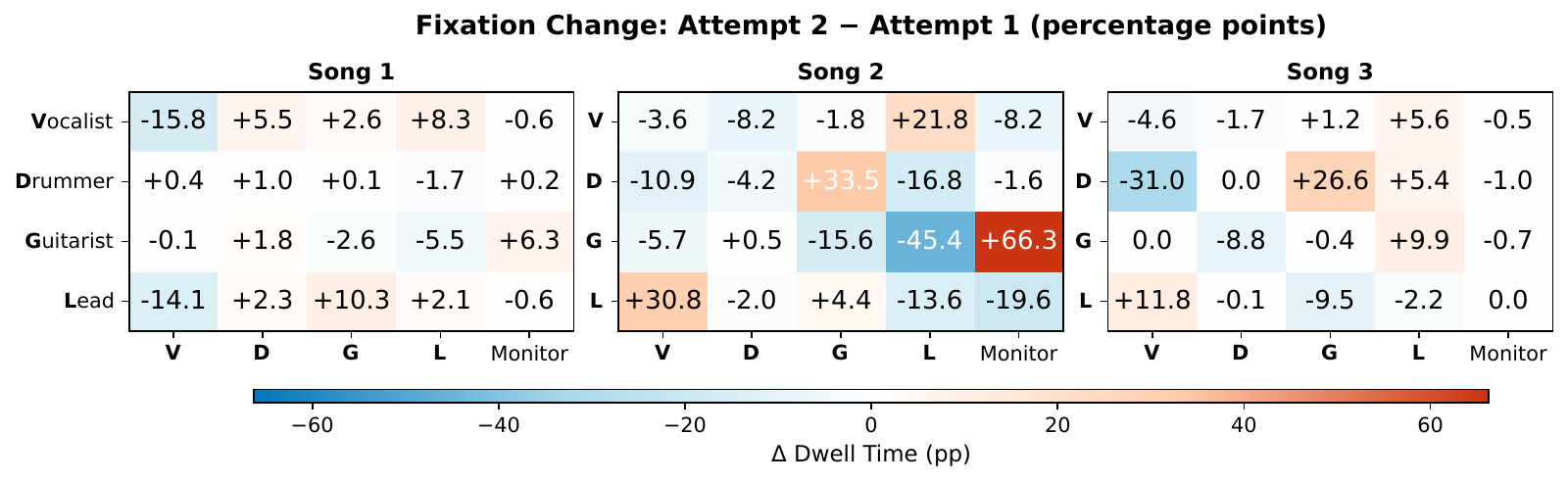}
        \caption{Dwell-time change (percentage points) from A1 to A2.}
        \label{fig:dwell_delta}
    \end{subfigure}
    \hfill
    \begin{subfigure}[t]{0.48\textwidth}
        \centering
        \includegraphics[width=\textwidth]{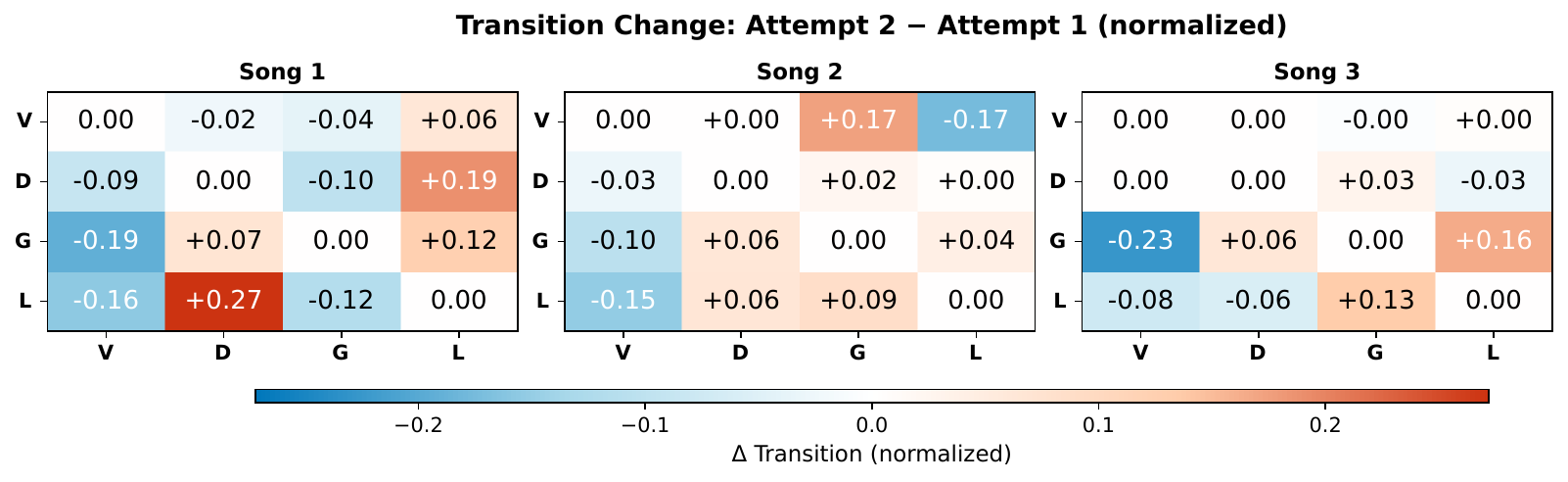}
        \caption{Transition count change from A1 to A2.}
        \label{fig:transition_delta}
    \end{subfigure}
    \caption{Changes between Attempt~1 and Attempt~2 for each song. S2 shows the largest redistribution, driven by the shift from teaching (A1) to performance (A2). S1 remains relatively stable across attempts. For~(a), blue cells indicate increased attention; red cells indicate decreased attention, and for~(b), blue cells indicate more transitions; red cells indicate fewer.}
    \Description{Two side-by-side panels, each containing three heatmaps (one per song). Left panel shows dwell-time deltas with a diverging blue-red colormap. Right panel shows transition count deltas with the same colormap. S2 panels show the most intense coloring, indicating large changes. S1 panels are mostly neutral, indicating stability.}
    \label{fig:delta}
\end{figure*}

Figure~\ref{fig:delta} shows dwell-time and transition changes from A1 to A2. S2 was the most volatile: during A1, the Lead stopped the group at the 4-minute mark, initiating eight minutes of structural teaching. In A2, the Guitarist redirected attention from the Lead (58.4\%$\to$13.0\%) to the monitor (17.4\%$\to$83.7\%), with transitions dropping 71\%---the teaching had equipped the Guitarist to use tablature independently. S3 contrasted sharply: the Lead's diagnostic feedback ("Your right hand is playing not in time") lacked structural scaffolding, deepening the Guitarist's attentional narrowing (96.2\% dwell, 96.9\% interpersonal, 41 transitions). This suggests structural teaching enables independence while diagnostic feedback alone reinforces leader-dependence. S1 (familiar song) remained stable (+5\% transitions) with converged gaze patterns.

\subsection{Temporal Gaze Patterns}

\begin{figure*}[t]
    \centering
    \includegraphics[width=\textwidth]{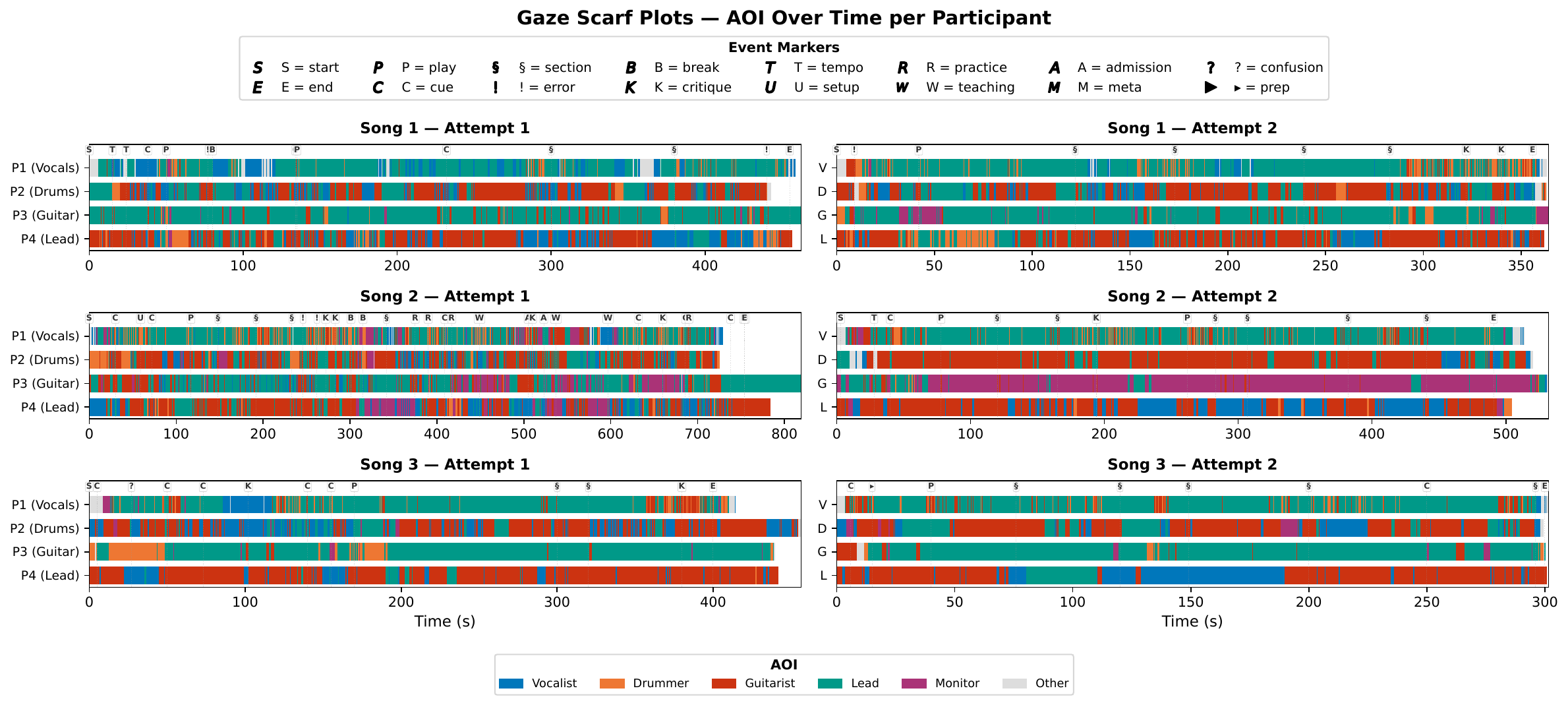}
    \caption{Scarf plots of sequential AOI fixation targets for all six sessions (3~songs~$\times$~2~attempts). Each row is one participant; colored segments indicate the fixation target, with grey for unclassified fixations. Vertical markers denote events (stops, teaching, count-ins). S2~A1 shows fragmented scanning with frequent stops; S3~A2 shows stable, sustained attention.}
    \Description{Multi-row timeline visualization showing all six sessions. Each row shows a horizontal bar divided into colored segments representing different AOI targets over time. Event markers appear as vertical lines. S2 A1 has many short, varied color segments and frequent event markers. S3 A2 has longer, more uniform segments and no event markers, indicating stable gaze.}
    \label{fig:scarf}
\end{figure*}

The scarf plots (Figure~\ref{fig:scarf}) provide a sequential view of each participant's gaze allocation over time, annotated with transcript events (count-ins, stops, teaching moments). Among the six sessions, S2~A1 and S3~A2 illustrate contrasting dynamics. In S2~A1, fragmented scanning clusters around stop events---when the Lead halts the group at 04:22, the Guitarist's gaze oscillates rapidly between Lead, monitor, and self-targets. In S3~A2, the Guitarist's timeline is dominated by long, unbroken Lead-directed segments, with no stop events. The Drummer shows steady alternation between targets throughout, fitting a timekeeper's scanning role---a contrast between rehearsal in crisis and rehearsal in flow.

\section{Discussion}

\paragraph{The Hub-and-Spoke Topology}

The hub-and-spoke structure is consistent with leader--follower dynamics in string quartets~\cite{wing2014optimal, davidson2005social}, extended here to a rehearsal context where the leader also teaches. Part of this concentration is positional (Figure~\ref{fig:teaser}(B)), yet the Guitarist's dwell reached 96.2\% (96.9\% interpersonal)---far exceeding positional accessibility---and the Drummer received only 5.5\% despite moderate visibility, pointing to functional, role-driven allocation. The most asymmetric relationship was the Vocalist$\to$Lead axis (61--85\% dwell vs.\ 10--28\% reciprocated, rising to 44\% in S2~A2), suggesting the Vocalist used the Lead as a timing reference without reciprocal monitoring; transcript evidence confirms corrections were directed almost exclusively at the Guitarist.

\paragraph{Attentional Narrowing as a Candidate Difficulty Marker}

The Guitarist's gaze pattern---up to 97\% interpersonal dwell on a single target with as few as 41 transitions---exhibits the hallmarks of attentional tunneling~\cite{easterbrook1959effect, wickens2009attentional}. However, extreme dwell does not track performance quality: the cleanest run (S3~A2: 96.2\% dwell, 41~transitions) and a confused session (S1~A1: 89.1\%, 128~transitions) show comparably high levels, suggesting two modes distinguishable by transition count. In \emph{anxious tunneling} (S1~A1), the performer locks onto a reference yet repeatedly seeks other anchors, as if uncertainty drives compulsive re-checking; in \emph{settled tunneling} (S3~A2), scanning effectively ceases from committed reliance on a single source. Feedback type reinforces this distinction: structural teaching in S2 equipped the Guitarist to use tablature independently (monitor dwell 17.4\%$\to$83.7\% in~A2), while diagnostic feedback in S3 deepened narrowing---scaffolding appears necessary to resolve leader-dependence, not merely identify it.

\paragraph{Participant Reflections}

Post-session review with the musicians confirmed several quantitative patterns. The Lead acknowledged his attentional bias in real time (``\textit{I'm just looking at [the Guitarist]\ldots{} proves our point}''), recognizing it as a byproduct of the teaching dynamic. The Guitarist's self-reports corroborate the narrowing interpretation: ``\textit{I don't know how to go back because I don't know your part.}'' This reveals compensatory watching without comprehension---a distinction important for pedagogy, as gaze fixation on an instructor does not necessarily mean the learner is extracting useful information. The Drummer's gaze revealed compensatory monitoring: in S2~A2, the Drummer directed 88.2\% of dwell to the Guitarist---more than any other member. When the Lead withdrew monitoring after teaching, the Drummer filled the gap, suggesting the group collectively maintained vigilance over the struggling member.

\paragraph{Self-Gaze and Instrument Ergonomics}

The large Other category (38\% mean dwell) likely subsumes substantial self-instrument monitoring ---particularly for the Drummer, whose kit demands precise stick placement across a wide visual arc, unlike the guitar neck (largely peripheral) or vocal performance (no instrument gaze required). These ergonomic differences likely contribute to role-level dwell asymmetries, and future work should separate ``own instrument'' from genuinely unclassified gaze within the Other category.

\paragraph{Implications for Music Practice}

Although preliminary, the contrast between S2 and S3 suggests that \emph{structural} teaching (explaining song form) enables learners to use references independently, while diagnostic corrections may reinforce leader-dependence---instructors could prioritise scaffolding when gaze data reveal narrowing. More broadly, transition rates and dwell distributions could serve as objective rehearsal reflection tools, helping identify coordination difficulties without relying solely on subjective assessment.

\section{Limitations}

With four participants and six sessions, all findings are descriptive; the study is a methodological pilot, not a confirmatory experiment. Songs were practiced in fixed order, confounding song-specific with session-order effects. The YOLO classifier's moderate performance (precision 0.75, recall 0.79, mAP@50 = 0.79) and the large ``Other'' category (38\% mean dwell) limit interpersonal resolution, and the small sample ($n=24$) limits generalizability despite the significant dwell--transition correlation ($r=-0.44$, $p=0.03$). We frame our main findings as testable hypotheses: (H1)~learners under difficulty exhibit attentional narrowing (high single-target dwell, low transitions), (H2)~structural teaching reduces leader-dependence more than diagnostic feedback, and (H3)~transition rates decrease with material familiarity beyond what shorter durations explain. Confirmatory work should use counterbalanced song orders, larger and more diverse ensembles, pre-registered analyses, rotated seating to control positional bias, and multiple genres.

\section{Conclusion}

We presented an exploratory analysis of visual attention during band rehearsal using mobile eye tracking and YOLO-based scene understanding. Our matrix-based approach reveals a hub-and-spoke topology with role-specific gaze signatures: the leader (broad scanning, high transitions), the learner (attentional narrowing, up to 97\% interpersonal dwell on one target), and the peripheral timekeeper (under-attended at 5.5\% mean dwell). Gaze transitions decreased up to 65\% on average (82\% individually) with material novelty, and feedback quality determines whether musicians shift from person-dependent to reference-dependent learning. These findings highlight the potential of matrix-based gaze visualizations for surfacing collaborative dynamics and generating hypotheses about pedagogical interventions such as narrowing detection and gaze-transition tracking. Future work should scale to larger ensembles, longitudinal designs, and real-time gaze feedback tools.

\section{Privacy and Ethics Statement}

All participants provided written informed consent for eye-tracking and scene camera recording, data are stored on encrypted drives with pseudonymized identifiers, and participants could withdraw at any time. Beyond this study, combining mobile eye tracking with automated person detection to infer cognitive states such as attentional tunneling raises privacy concerns if repurposed outside consensual rehearsal contexts---for instance, to surveil or evaluate musicians without their knowledge---and we caution that any deployment of this pipeline should carry explicit participant awareness and institutional oversight.

\bibliographystyle{ACM-Reference-Format}
\bibliography{main}

\end{document}